\documentclass[twocolumn]{article}
\usepackage{authblk}
\usepackage{amsmath,amsfonts,bm} 

\usepackage{xcolor}
\usepackage{color}
\usepackage{graphicx}
\usepackage{graphics}
\usepackage{amsmath}
\usepackage{amssymb}
\usepackage{amsfonts}
\usepackage{latexsym}
\usepackage{wasysym}
\usepackage{mathrsfs}
\usepackage{comment}
\usepackage{amssymb} 
\usepackage{cleveref}
\providecommand{\keywords}[1]
{
	\small	
	\textbf{\textit{Keywords---}} #1
}

\renewcommand\d\delta
\newcommand\D\Delta

\definecolor{light-blue}{rgb}{0.8,0.85,1}

\newcommand\beq{\begin{equation}}
	\newcommand\beqn{\begin{eqnarray}}
		
		\newcommand\eeq{\end{equation}}
	\newcommand\eeqn{\end{eqnarray}}

\newcommand{\DE}{\mathrm{d}}

\newcommand{\VE}{\mathbf}

\newcommand{\der}[2]{\frac{\DE#1}{\DE#2}}

\newcommand{\pder}[2]{\frac{\partial#1}{\partial#2}}

\newcommand{\integ}[3]{\int_{#1}\!{#2}\,\DE #3}

\newcommand{\oointeg}[3]{\oint_{#1} \!{#2}\,\DE #3}
\newcommand{\eqrf}[1]{Eq.{\labelcref{#1}}}
\newcommand{\lamax}{\lambda_{\text{max}}}
\newcommand{\lamin}{\lambda_{\text{min}}}
\newcommand{\laco}{\lambda_{c0}}

\renewcommand\d\delta

\definecolor{light-blue}{rgb}{0.8,0.85,1}

		\title{A predictive microstructure-based approach for the anisotropic damage, residual stretches and hysteresis in biodegradable sutures}

\author[1]{Gennaro Vitucci}
\author[1]{Domenico De Tommasi}

\author[1]{Giuseppe Puglisi}
\author[1]{Francesco Trentadue \footnote{Corresponding author: francesco.trentadue@poliba.it }}

\affil[1]{DICATECh, Politecnico di Bari, Via Re David 200, 70125 Bari, Italy}

\date{}

\begin{document}
\maketitle

	\begin{abstract}
		We propose a predictive model for the mechanical behavior of biodegradable polymers of interest for biomedical applications. Starting from a detailed description of the network behavior of the copolymer material, taking care of bonds breaking and recrosslinking effects, folded-unfolded transitions and network topological constraints, we deduce a macroscopic  law for the complex mechanical behavior of many biomedical materials with a particular focus on absorbable suture threads and a perspective to new material design. A crucial novelty of the model is the careful description of the observed microscopic anisotropic damage induced by the deformation, here described based on the classical microsphere integration approach. The resulting energy, characterized by a few number of material parameters, with physically clear interpretation, is successfully adopted to predict our cyclic experiments on poliglecaprone sutures with anisotropic damage, permanent stretch and internal hysteresis. By also studying other suture materials behavior we show that the predictivity properties of the model can be extended also to materials with a different mechanical response and thus also possibly applied for the design of new, high-performance biomedical materials.
		
	\end{abstract}

\section{Introduction}
Biodegradable copolymers are successfully adopted in different biomedical applications.
In particular synthetic polymers such as polylactic acid (PLA), polyglycolic acid (PGA), and polycaprolactone (PCL) have been adopted for their biocompatibility and biodegradability.
With specific reference to suture applications, beyond the quoted properties, also the mechanical response of the material under cyclic loading is fundamental in the everyday medical practice. The prediction of residual stretches, toughness, stress softening, good tensile and knot-pull strengths are crucial in clinical applications. The effective microstructure characterization of these materials is then key both for predicting the behavior of existing suture threads and for the design of new engineering materials with prescribed macroscopic material response.

In this paper, we extend our previous 
approach proposed in  \cite{trentadue2021predictive} by accounting for a fully three-dimensional point-wise distribution of polymeric chains whose properties vary with stretch history. As well known, soft materials are characterized by a fundamental anisotropy of damage since each chain natural configuration depends on its own extension history. 
Specifically, in the framework of statistical mechanics, we propose a constitutive approach, in which the Helmholtz free energy is composed of two terms: an entropic Worm Like Chain (WLC) component
and a network topological one. An important novelty of the WLC term is that, to consider the important effect of unloading, we assume, again based on statistical mechanics results, 
that both the natural and contour length of each chain depend on molecule stretch histories. In particular we describe the copolymer molecules as a two phase material with hard (folded) and soft (amorphous) domains. 
Due to externally imposed stretching, folded domains undergo a hard$\rightarrow$soft transition with an increase of available monomers producing a contemporary growth of the contour and natural lengths. These processes typically involve exchange of heat with the environment due to formation and dissociation bond enthalpy and as such should rigorously be viewed as complex multiphysics phenomena \cite{rey2013influence,martinez2013mechanisms}. For the sake of simplicity and in order to deal only with purely mechanical quantities, we instead describe bonds rupture and recrosslinking and the resulting microscopic dissipation in an isothermal setting. By following the approach  proposed in \cite{doi1988theory,goktepe2005micro} we then introduce a second important energy term, taking care of the network topological constraint. Based on the deduced filament Helmholtz free energy, we consider, by following the classical microsphere approach \cite{bavzant1986efficient,goktepe2005micro},  an isotropic chains distribution of the copolymer. As a result, for an assigned deformation history, we are able to deduce the chain stretch in each direction that, after numerical integration, 
gives back the total internal specific energy with history dependent parameters, describing the anisotropic damage effect. The resulting macroscopic model is characterized by a set of five parameters only: one for each component of the free energy, one relating the contour length to the natural end-to-end chain length and two describing refolding and unfolding in unloading and reloading. In this way we are able to capture both anisotropic damage, residual stretches, and hysteresis. 

As a consequence, a micromechanically-based model for biodegradable isotropic polymers is developed and validated with experimental data on biodegradable copolymer suture threads. The comparison with our experiments on poliglecaprone 25 (Monocryl\textsuperscript{\textregistered} \cite{bezwada1995monocryl}) shows the excellent capacity of the model in predicting the experimental behavior. To further support the proposed model, in \ref{app:zu} we  successfully reproduce the cyclic behavior of different absorbable sutures by introducing a further parameter for the internal hysteresis. The resulting model exhibits significant improvements, as compared with previous models in the literature \cite{elias2013stress,elias2014rule} in the prediction of the sutures experimental behavior under cyclic loading.
\section{Methods}
\begin{figure}[b!]
	\centering
	\includegraphics[width=.40\textwidth]{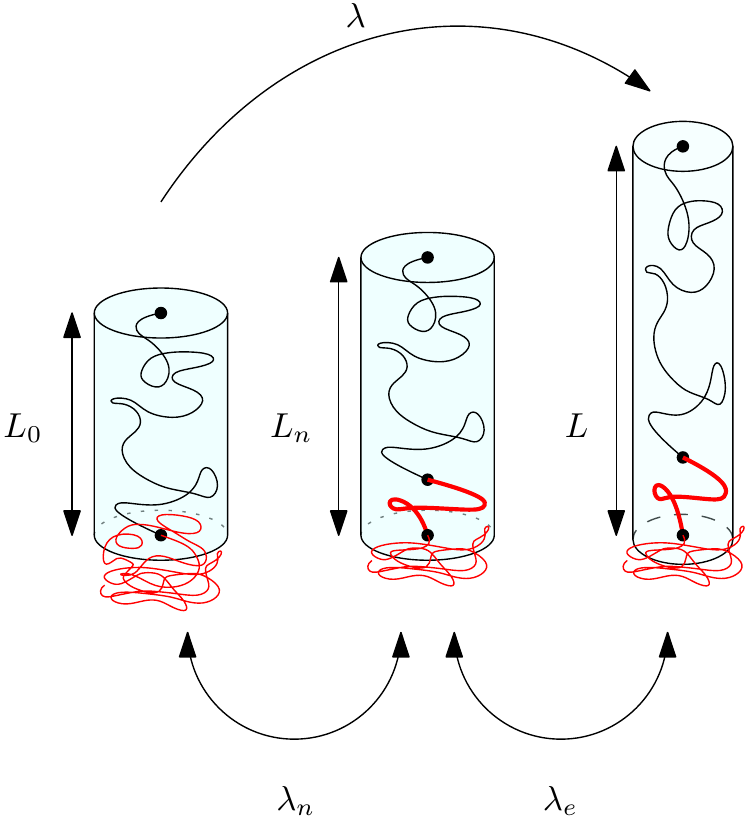}
	\caption{Cartoon of the chain deformation $\lambda$ decomposition into an elastic and a natural stretch $\lambda_e$ and $\lambda_{n}$, respectively. The light blue tube represents the topological constraint arising from the chain interaction within the network. The amorphous region length (thin red line outside the tube) does not contribute to the filament length $L$ since it is expected to be small compared to the unfolded region (hidden length).}
	\label{fig:chstr}
\end{figure}

\begin{figure}[ht!]
	\centering
	\includegraphics[width=.40\textwidth]{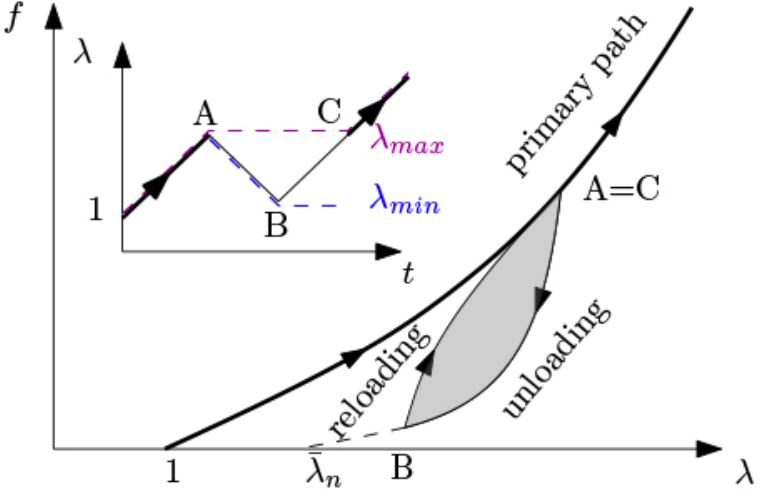}
	\caption{Filament loading history. Inset) Stretch vs time. The magenta line shows the primary path, the path AB unloading, the path BC reloading (see \eqrf{eq:cases}). Main) Force vs stretch. An internal hysteresis cycle (gray) is produced along the path ABC. The residual stretch $\bar{\lambda}_n$ is the stretch at zero force and evolves with the primary path. }
	\label{fig:fstr}
\end{figure}

In this section we extend a model for biodegradable sutures proposed by some authors of this paper \cite{trentadue2021predictive} in several directions. On one side we consider the important anisotropic effect in the evolution of the natural molecules configurations and the topological network constrains, on the other side, by introducing recrosslinking effect, we describe the important experimental effect of internal hysteresis cycles. Similarly we describe the insurgence of residual stretches fundamental for modelling the clinical application of the suture. 

Our model starts from the observation of the behavior at the molecular scale. While our previous model was based on a simplifying assumption of non-interacting molecules in the direction of the suture, here we consider a more detailed analysis considering of a three-dimensional network of chains undergoing cyclic uniaxial stress histories. 

As in usually adopted schemes of rubber elasticity, our multiscale model is based on the {\it affinity assumption} identifying macroscale stretches with the molecular ones \cite{rubinstein2003polymer}. In particular our approach is based on the microsphere integration scheme \cite{bavzant1986efficient}, firstly applied by \cite{goktepe2005micro,miehe2004micro,miehe2005micro} to model polymer networks mechanical behavior. In this way, as we show in the following, we are able to describe in detail the evolution of anisotropic damage and residual stretches.

\begin{figure}[ht!]
	\centering
	\includegraphics[width=.9\columnwidth]{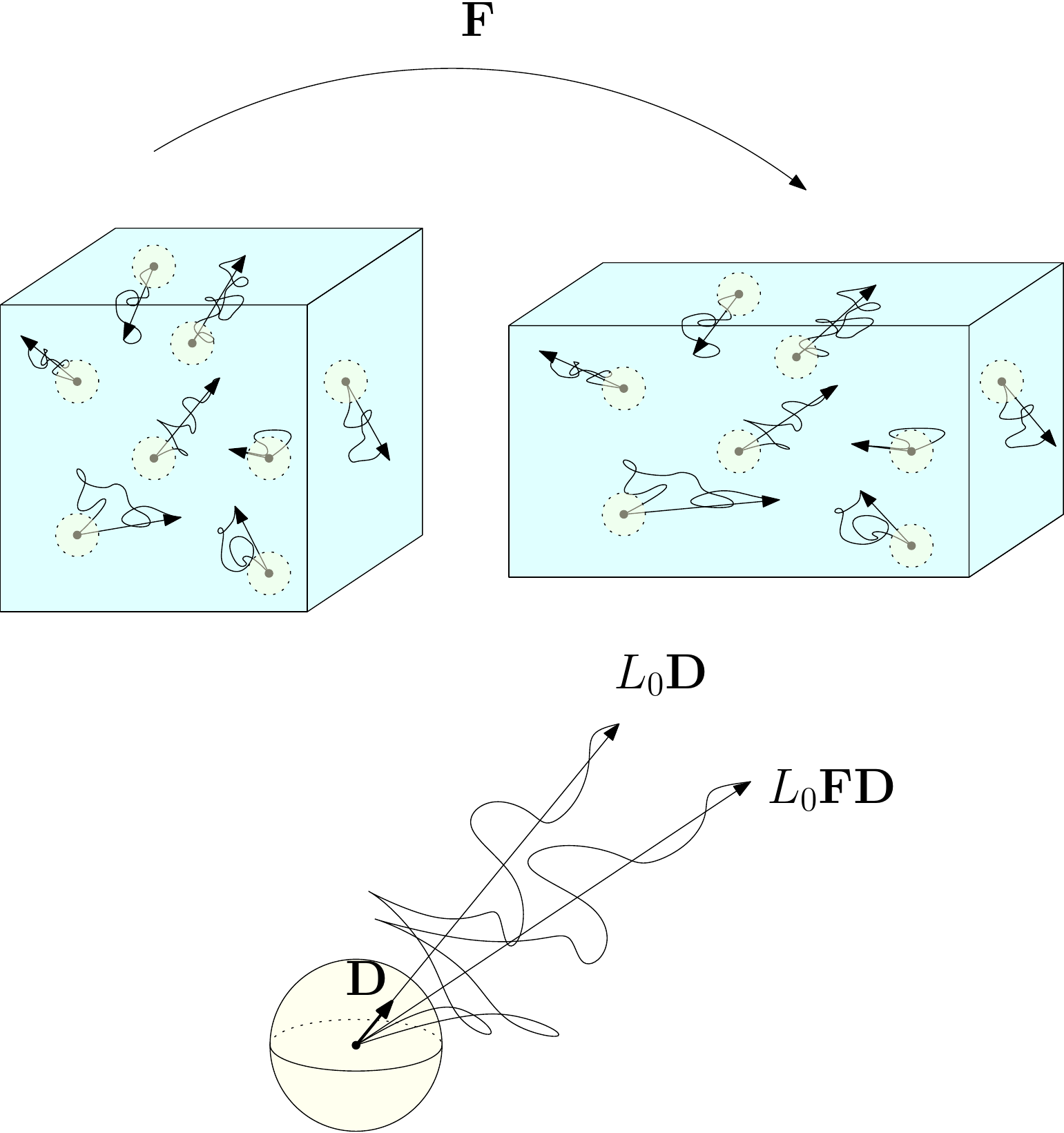}
	\caption{Micro-macro scale transition. The representative volume (light blue) contains a distribution of filaments which are isotropically oriented in space along the unit vector $\VE{D}$ whose directional cosines are measured on the microsphere (yellow). The macroscopic deformation gradient $\VE{F}$ stretches and rotates the filaments according to the affinity hypothesis.}
	\label{fig:mM}
\end{figure}

Our microstucture scheme is represented in Fig.\ref{fig:mM}. As described in \cite{de2013energetic,de2015multiscale,puglisi2017mic,de2010damage} copolymers molecules are constituted by alternating sequences of coiled and folded domains, here interpreted as two different material phases. We then deduce  the macroscale material behavior as an homogenized 
effect of complex links breaking and recrosslinking \cite{qi2005stress,yeh2003situ}. Roughly speaking, 
we can imagine that as we apply a given deformation the hard domains of elongated chains undergo unfolding, corresponding from one side to a hard$\rightarrow$soft material transition, from the other side to a variation of the molecule expected length, due to the availability of a larger number of free monomers. When unloaded the chains refold to simple entropic effects and consequently some broken links can rebind. 

As described for instance in \cite{marckmann2002theory,dorfmann2004constitutive,goktepe2005micro,diani2006damage,machado2012induced} these effects are strongly anisotropic, so that an effective description of the material behavior requires accounting for different elongations in different directions and this is at the base of the proposed predictive model for suture materials.

\subsection{Single chain}

Following the approach in \cite{trentadue2021predictive}, we base our model on Flory's assumption that the expected end-to-end length $L_n$ of the co-polymer macromolecules composed of $n$ Kuhn segments of length $b$ can be estimated, based on classical statistical mechanics (see \cite{kuhn1942beziehungen}), as
\begin{equation}\label{eq:florat}
	L_n^2=C_\infty nb^2,
\end{equation} 
where Flory's characteristic ratio $C_\infty$ is unitary for ideal Gaussian chains, whereas it grows with the persistence length \cite{rubinstein2003polymer}. 

To apply the recalled affinity assumption we may introduce the following stretch of the macromolecules:
\begin{equation}
	\lambda=\frac{L}{L_{0}},\quad \lambda_n=\frac{L_n}{L_{0}},\quad\lambda_c=\frac{L_c}{L_{0}}\quad
\end{equation}
where $L_0$ is the expected length of the previously unloaded chain, whereas $L_c=n b$ is the contour length, measuring the limit chain extensibility.
Interestingly, by using \eqrf{eq:florat}, we may observe \cite{trentadue2021predictive} that
the contour stretch $\lambda_c$ is related to the natural one $\lambda_n$ via the initial contour stretches $\lambda_{co}$:
\begin{equation}\label{eq:lanlac}
	\lambda_c= \lambda_{co}\lambda^2_n,\qquad\lambda_{co}=\frac{L_0}{C_\infty b}.
\end{equation}
Observe that, in accordance with classical finite elasto-plasticity, the total stretch $\lambda$ is multiplicatively decomposed (see Fig.\ref{fig:chstr}) into an elastic and a permanent part
\begin{equation}\label{eq:lelast}
	\lambda=\lambda_e\lambda_n.
\end{equation}
As a result, the model has a single internal variable describing inelastic behavior that we choose as the natural stretch $\lambda_n$. Notice that we neglect the length of the folded domains thus assuming the existence of the so-called \textit{hidden length} that  is involved in the model due to the rupture of sacrificial bonds during unraveling as commonly understood in biopolymers mechanics \cite{fantner2006sacrificial}.

In this work we consider for the chain the following Helmholtz free energy density (per unit contour length)
\begin{equation}\label{eq:freeen}
	\psi(\lambda,\lambda_n)=\kappa\dfrac{\left(\lambda-\lambda_n\right)^2}{\lambda_c -\lambda}-\nu \lambda\left(1+\lambda_e^{-2}\right),
\end{equation}
where the first term describes the worm-like chain energy type used in \cite{trentadue2021predictive} whereas the second one has been proposed in \cite{de2013energetic,marko1995stretching} and captures a topological constraint arising from the chain entanglement with its environment and itself. This term is expected to be small for semiflexible chains. Loosely speaking, this term takes care of a Poisson type confinement effect for a chain elongating inside the surrounding network.The proportionality with $\lambda_e^{-2}$ is linked to the aerial deformation. Indeed, at given $\lambda_n$ this energy terms decreases as $\lambda_e$ grows to describe a decreasing entropic term that may be loosely interpreted as a lower energy required to elongate chains constrained in thinner tubes.

The first principle of thermodynamics for our isothermal system gives
\begin{equation}\label{eq:1therm}
	f\dot{\lambda}(t)=\dot{\psi}(\lambda_n(t),\lambda(t))+\gamma(t),
\end{equation}
where $\gamma$ is the rate of energy dissipation and $t$ is a time type variable describing the loading history in the proposed rate-independent behavior. Following the classical settings of \cite{col67} for thermodynamical systems with internal state variables, we find the equilibrium force $f$ as
\begin{equation}\label{eq:fpsi}
	\begin{split}
		f&=\pder{\psi}{\lambda}=f_\kappa+f_\nu\\
		&=\kappa\left[\left(\frac{\lambda_c-\lambda_n}{\lambda_c-\lambda}\right)^2-1\right]-\nu\left[1-\lambda_e^{-2}\right].
	\end{split}
\end{equation}

As anticipated, here we consider the possibility that, as in the case of knots formation and/or other clinical conditions, the suture can undergo cyclic force loading as represented in Fig.\ref{fig:fstr}. Next, we assume the following phenomenological evolution of $\lambda_n$ depending on loading history, which can be distinguished into the following conditions (see Fig.\ref{fig:chstr}):
\begin{equation}\label{eq:cases}
	\left\{\begin{array}{llll}
		\lambda=\lamax,& \dot{\lambda}>0&...&\text{primary path}\\
		\lambda=\lamin,& \dot{\lambda}<0&...&\text{unloading path}\\
		\lamin<\lambda<\lamax,& \dot{\lambda}>0&...&\text{reloading path}\\
	\end{array}\right .
\end{equation}
To attain an explicit analytic description of the complex breaking and recrosslinking phenomena, we consider special memory properties of the material. Specifically, we assume that on the primary path the macromolecule unfolds depending on the maximum attained stretch $\lambda_{max}$. On the other hand, as anticipated, we assume that under unloading there is a partial rebinding and under reloading a new unbinding of reformed links. Furthermore the memory of the minimum stretch $\lamin$ is reset at every cycle. 

The evolution of breaking and recrosslinking effects during loading, unloading and reloading depends on the specific properties of the chains and of the network. Here, based on an energetic interpretation proposed in \cite{de2013energetic}, we assume the following evolution law for the internal variable
\begin{equation}\label{eq:lacevol}	
	\lambda_n=\sqrt{\lamax}\left(\frac{\lambda}{\lamax}\right)^{{\beta_u+\beta_r\sqrt{\lambda-\lamin}}}\geq\bar{\lambda}_n\quad
\end{equation}
depending on two only positive scalar material parameters $\beta_u$ and $\beta_r$ measuring the unfolding and refolding evolution. 

In particular, during primary loading we get 
\begin{equation}\lambda_n=\sqrt{\lamax}, \quad\lambda_c=\lambda_{co} \lambda_{max}
	\label{eq:lc}\end{equation} that can be obtained using 
\eqrf{eq:lanlac} under the assumption that $L_c\propto L_{max}$ descending by the hypothesis of a constant dissipation rate per unit elongation.
Notice that in view of \eqrf{eq:cases,eq:lacevol} we have a purely reversible elastic response in compressed fibers. 

The second factor in \eqrf{eq:lacevol} accounts for internal hysteresis effects.
Specifically, the zero-force residual stretch $\bar{\lambda}_n$ of \eqrf{eq:lacevol} is determined by imposing the condition $\lambda=\lambda_n$, which gives
\begin{equation}\label{eq:lastar}
	\bar{\lambda}_n=\lamax{}^{ \frac{1}{2}\frac{1-2\beta_u}{1-\beta_u} },
\end{equation}
i.e. the residual stretch depends on the maximum historic stretch in the particularly simple form of a power law.
This prospects the possibility of experimentally determining the exponential factor $\beta_u$ and moreover assigns also the variation of the contour stretch $\bar{\lambda}_c=\lambda_{co}\bar{\lambda}_n^2 $. 

To continue with a microstructure interpretation of \eqrf{eq:lacevol}, during an internal cycle
${\lambda}_{min}(=\bar{\lambda}_n) \leq \lambda< \lambda_{max}$ as in Fig.\ref{fig:fstr},  there exists	a fraction $\rho$ of reversibly unfolded Kuhn segments with respect to the cumulative number of unfolded monomers. Specifically, since the contour stretch is proportional to the total number of unfolded segments $n$, via \eqrf{eq:lanlac}, \eqrf{eq:lastar}, and \eqrf{eq:lc} we obtain 
\begin{equation}
	\rho(\lamax)=\frac{\hat{\lambda}_c(\lamax)-\bar{\lambda}_c}{\hat{\lambda}_c(\lamax)}=1- \lamax {}^{-\frac{\beta_u}{1-\beta_u}},
\end{equation}   
where the function $\lambda_c=\hat{\lambda}_c(\lamax)$ is determined by \eqrf{eq:lanlac} and \eqrf{eq:lacevol}.  
Observe that the recovering increases with $\lambda_{max}$ and this will be confirmed by the following experimental analysis. Moreover, after reloading, when the stretch reaches again its previously attained maximum value $\lambda=\lamax$, these recovered fraction is again unfolded. In this way the internal hysteresis equals the unfolding energy of this fraction. Eventually, the system memory respects the Return Point Memory property \cite{bertotti1998hysteresis}, because when $\lambda=\lamax$ the system goes back exactly to the starting configuration in the primary loading path.

Thus, the two additional parameters $\beta_u$ and $\beta_r$ describe internal hysteresis. In particular $\beta_u$ measures refolding properties during unloading, thus influencing also the unfolding variation of natural length, whereas $\beta_r$ measures the hysteresis size due to refolding during reloading (\cite{de2013energetic,de2015multiscale,puglisi2017mic}). All things considered,
our single chain model is able to describe the damage, the hysteresis and the residual stretches in loading unloading and reloading via only five parameters. 

\paragraph{Thermodynamic consistency}
We recall that our modeling is inscribed in the framework of thermodynamics with internal state variables in the special simple
setting of \textit{isothermal} processes \cite{col67}, so that we do not consider the contribute of the heat flux to the dissipation rate.    
From \eqrf{eq:1therm} and \eqrf{eq:fpsi} one obtains that the rate of dissipation $\gamma(t)$ of a single fiber in the particular case of an isothermal process is
\begin{equation}\label{eq:gammageq0}
	\gamma(t)=-\pder{\psi}{\lambda_n}\dot{\lambda}_n\geq 0\; .
\end{equation}
In view of \eqrf{eq:freeen} and \eqrf{eq:lanlac} it reads
\begin{equation}
	\begin{split}
		\gamma(t)=
		\left( 2 \kappa\dfrac{ \lambda_e (\lambda-\lambda_n) }{\lambda_c-\lambda} + 2 \dfrac{\nu}{\lambda_e}\right) \dot{\lambda}_n\ge 0
	\end{split} \; .\label{eq:gammageq1}
\end{equation}
For fibers in compression $(\lambda<\lambda_r)  $ we have considered a purely elastic response, then  $ \dot{\lambda}_n=0 $ and \eqrf{eq:gammageq1} is clearly satisfied. For fibers in traction $(\lambda>\lambda_r)$ we always have 
\begin{equation}\label{eq:labounds}
	\lambda_n\leq\lambda<\lambda_c	,
\end{equation}
and then \eqref{eq:gammageq0} is satisfied if the elastic constants $ \kappa $   and $ \nu $ are positive and $ \dot{\lambda}_n> 0 $, whereas is not satisfied in an unloading path. However, as typical for hysteresis cycles, the global condition
\begin{eqnarray}\label{eq:disstot}
	\oointeg{\text{cycle}}{f(\lambda)\dot{\lambda}}{t}\geq 0,
\end{eqnarray}
granting dissipative internal cycles, is numerically verified and regulated by the parameter $\beta_r$.
We may interpret this result as due to the physical condition that part of the heat production accompanying links scissions can be reused by the system to partial healing and refolding, with a global thermodynamic consistency of each cycle.

\subsection{Network model}
To model the macroscopic thread behavior, we assume that it is composed by a network of isotropically oriented chains in the framework of rubber elasticity \cite{goktepe2005micro}. 
Furthermore, by the classical affinity assumption, we correlate the chain deformation with the macroscopic one. Thus, if $\VE{F}$ is the macroscopic deformation gradient, a (unitary length) chain oriented along $\VE{D}$ in the reference configuration is transformed in the present configuration into the deformed vector
\begin{equation}
	\VE{d}=\VE{F}\VE{D},\qquad \lambda=\|\VE{d}\|.
\end{equation}
Under the considered isotropic assumption, the probability that a chain points in the direction associated to  $ \VE{D} $ is $1/4\pi$ and then the mean total free energy of the microstructure is 
\begin{equation}\label{eq:Psi_mic}
	\Psi=\frac{1}{4\pi}\integ{\text{sphere}}{\psi(\lambda,\lambda_n)}{\Omega} \; . 
\end{equation}
As in \cite{col67}, one can obtain the first Piola-Kirchhoff stress tensor $\VE{\Sigma}$ as 
\begin{equation}\label{eq:genpsi}
	\VE{\Sigma}=\pder{\Psi}{\VE{F}}.	
\end{equation}

\paragraph{Macroscopic uniaxial extension}
While our model can be applied to generic deformation classes, such as in the case of biological tissues, in this paper we focus on the  case of uniaxial extension in the thread direction, because this is the case of interest in biomedical applications.

Thus, under the classical hypothesis of incompressible materials, i.e. $\det\VE{F}=1$, the deformation gradient in a principal reference frame takes the form
\begin{equation}
	\VE{F}=\left[\begin{matrix}
		\Lambda & 0  & 0 \\
		0& \Lambda^{-\frac{1}{2}} &0 \\
		0& 0 & \Lambda^{-\frac{1}{2}}\\
	\end{matrix}\right].	
\end{equation}
Since the work on the system is performed by the only non-zero component $\sigma$ of $\VE{\Sigma}$, i.e. the normal traction, similarly to \eqrf{eq:genpsi} it results
\begin{equation}
	\sigma=\pder{\Psi}{\Lambda}=\frac{1}{4\pi}\integ{\text{sphere}}{\pder{\psi}{\lambda}\der{\lambda}{\Lambda}}{\Omega}.
\end{equation}
Integrals on the microsphere of the type \eqref{eq:Psi_mic} can be computed via the numerical scheme firstly proposed by \cite{bavzant1986efficient} 
\begin{equation}\label{eq:int_mic}
	\begin{split}
		\langle v \rangle=\frac{1}{4\pi}\integ{\text{sphere}}{v(\VE{D})}{\Omega}\approx\sum_{i=1}^{N}w_i v_i, \\
		\quad v_i=v(\VE{D}_i),\quad \sum_{i=1}^{N}w_i=1,
	\end{split}	
\end{equation}
where $w_i$ are weights and $N$ is the number of directions in which the sphere is numerically decomposed (in this work we use a number of directions $N=33$).
Thus we have
\begin{equation}\label{eq:sig_mic}
	\sigma\approx\sum_{i=1}^{N}w_i\der{\psi_i}{\Lambda}=\sum_{i=1}^{N}w_i f(\lambda_{i})\der{\lambda_{D_i}}{\Lambda},
\end{equation}
where the function $f(\lambda_{i})$ has been given in \eqrf{eq:fpsi}. 

Note that in our isochoric case we directly derive $\sigma$ from a mechanical energy balance where  pressure does not appear. This procedure is computationally cheaper than the general one (see \cite{goktepe2005micro,miehe2004micro,miehe2005micro}), where integration of microstresses is needed in order to compute all the nonzero components of the macrostress.

\section{Model validation and discussion}

Observe that very different mechanical behaviors can be attained regarding hardening, permanent stretches, and hysteresis. 
All these properties can be crucial in biomedical applications and since our parameters have a clear microstructure interpretation we believe that our model can also help in the direction of designing new bioinspired materials.

\begin{figure}[t!]\centering
	\includegraphics[width=\columnwidth]{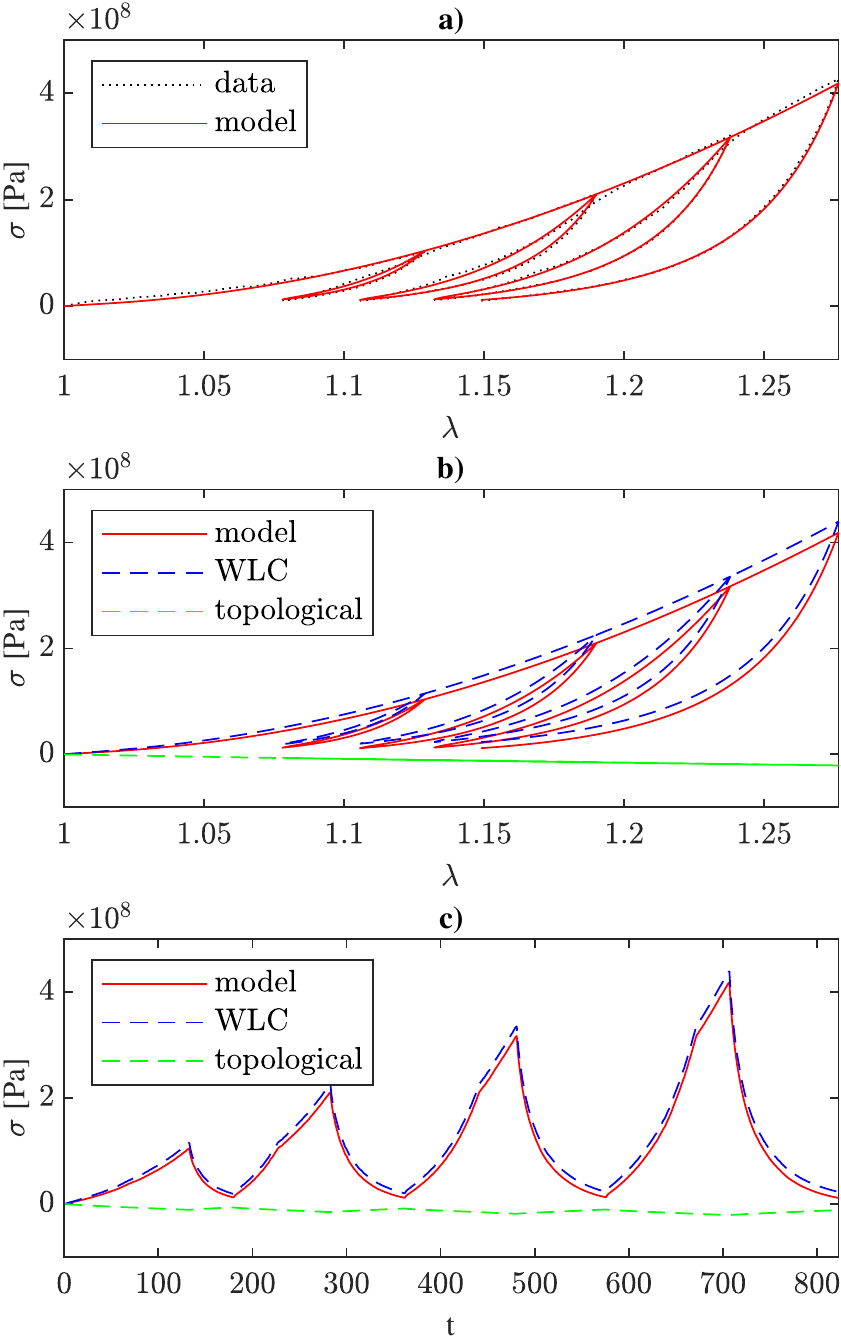} 
	\vspace{-.15 cm}\caption{Comparison of the experimental and theoretical stress-strain diagrams. Our experimental data, model parameters as in Table \ref{tab:pfit}. a) Overall performance of the current model. b) Separate contributions of WLC and topological stresses. c) Same as b) plotted as a function of the time order parameter $t$ highlighting that the topological stress gives maximum relative contribution toward low stresses.} 
	\label{fig:stresslambda}
\end{figure}

In the following, to validate the proposed model in capturing the experimental behavior of sutures, with complex cyclic loading and taking care of damage (Mullins effect) and residual stretches, together with internal hysteresis we refer to our experiments, previously described in \cite{trentadue2021predictive} by the simpler model recalled in the introduction, on 
diffuse synthetic absorbable Monocryl\textsuperscript{\textregistered} (poliglecaprone 25) monofilament sutures threads (see Fig.\ref{fig:stresslambda}a).

The vector of unknown constitutive parameters  $\mathbf{v} =(\kappa, \mu, \laco, \beta_u, \beta_r)$  is determined by minimizing the error function:
\begin{equation}
	\mbox{Err}(\mathbf{v})=\dfrac{| \boldsymbol{\sigma}_{exp}- \boldsymbol{\sigma}(\mathbf{v})|\,}{| \boldsymbol{\sigma}_{exp}|}\, .
	\label{err}
\end{equation}
Here  $ \left| \cdot \right|  $ is the 1-norm, $ \boldsymbol{\sigma}_{exp}$ is the extracted sample of experimental values of stress and  $ \boldsymbol{\sigma}(\mathbf{v})$ is the vector of corresponding analytic values computed as in \eqrf{eq:sig_mic}.
Next, the following optimization problem is posed
\begin{equation}
	\begin{cases}
		\min \mbox{Err}(\mathbf{v})\\
		\mathbf{v}^{l}\le\mathbf{v}\le \mathbf{v}^{u}\\
	\end{cases}\label{optprob}
\end{equation}
where
$\mathbf{v}^{l}$ and $\mathbf{v}^{u}$ are lower and upper bounds for the constitutive parameters, derived from physical considerations.
A global optimization procedure has been adopted, using the \textit{particleswarm} toolbox of MATLAB\textsuperscript{\textregistered} and adopting a population size of 400 elements. The obtained optimal values of the constitutive parameters are reported in Tab.\ref{tab:pfit}.

\begin{center}
	\begin{tiny}		\centering
		\begin{table}[h!]
			\small
			\begin{tabular}{ | c| c| c| c| c|  }
				\hline 
				$\kappa$[$10^8\times$Pa]&$\nu$[$10^8\times$Pa]&$\laco$&$\beta_{u}$&$\beta_r$\\
				\hline  
				0.148  &  3.548 &   1.008 &    0.825   & 0.359  \\
				\hline
			\end{tabular}\caption{Parameter identification for Monocryl. Error of the optimization procedure 2.62$\%$.}
			\label{tab:pfit}
		\end{table}
	\end{tiny}
\end{center} 
In Fig.\ref{fig:stresslambda}a 
a comparison between our model and the experimental data is shown. The error determined by \eqrf{err} is $\mbox{Err}(\mathbf{v}_{opt})= 2.62\%$, mainly regarding an experimental non perfect respect of the Return Point Memory assumption. Interestingly, despite the complication due to the introduction of internal hysteresis, we improved our previous results \cite{trentadue2021predictive} at fixed number of parameters thanks to both the adoption of the three-dimensional distribution of chains and the introduction of the topological term.

Specifically, in Fig.\ref{fig:stresslambda}b we decomposed the contribution of the two additive parts of the WLC chains $ \sigma_\kappa $ and of the topological constraint $ \sigma_\nu $,
where $t$ is a pseudo time parameter. It can be noted that, even if  the elastic constant $\nu $ is much greater than $\kappa$,  $ \sigma_\nu $ is much smaller than $ \sigma_\kappa $ and is always negative in accordance with our previously introduced physical interpretation. In particular the analysis of the numerical results suggests that this term plays a more important role in the description of the unloading states of the material. Moreover, it is evident from this numerical analysis that the internal hysteresis dissipation quantified as in \eqrf{eq:disstot} is produced chiefly by the WLC part of the stress, as expected from the fact that the topological one only contains the elastic stretch $\lambda_e$ (see Fig.\ref{fig:stresslambda}.c, green curve).

\begin{figure*}[ht!]
\centering
\includegraphics[width=.95\textwidth]{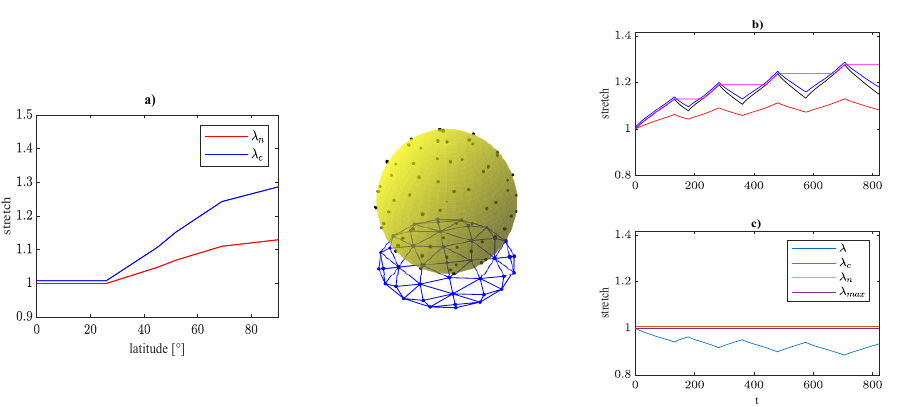} \label{fig:allfibersb}
\caption{Damage induced anisotropy. Microstructure uniaxially loaded along the north-south direction. Points on the microsphere and related projection onto the horizontal plane (blue, middle panel) show the $2\times 33$ direction cosines used for integration. a) Maximum damage per latitude on the microsphere. The filaments whose end-to-end unit vector is tilted less than $\approx30^\circ$ with respect to the horizontal result undamaged, since their $\lambda_{n}$ and $\lambda_c$ always stay at their initial value, i.e. one and $\laco$, respectively. b) and c) Evolution of the stretches for chains parallel and perpendicular to uniaxial load.}
\end{figure*}	
Finally, we intend to show how anisotropy is induced in an initially isotropic material by means of differential damage in different directions. To this aim, in Fig.\ref{fig:allfibersb}a we show the distribution of the permanent set $\lambda_c$ and $\lambda_{n}$  with respect to latitude.
Filaments with a latitude smaller than $\approx30^\circ$ with respect to the horizontal result undamaged, i.e. their $\lambda_{n}$ and $\lambda_c$ always stay at their undamaged value. In particular Fig.\ref{fig:allfibersb}b and Fig.\ref{fig:allfibersb}c refer to chains parallel and orthogonal to the uniaxial load, respectively. The first ones accumulate damage during growing cyclic loading, whereas the second ones result unaffected, since $\lamax$ is never attained. 

Finally to describe the effectiveness of the model in describing the experimental behavior of different suture materials, in the 
Appendix we successfully predict the experimental behavior of sutures reported in  \cite{elias2013stress}.

\vspace{0.6 cm}

\section{Conclusions}
We propose a predictive model for suture materials. The constitutive assumptions extend our previous model in \cite{trentadue2021predictive}, where a WLC energy term introducing variable natural and contour lengths was proposed.  As well known (e.g. see \cite{puglisi2017mic}), soft macromolecular materials undergo anisotropic damage. This was already noticed in \cite{mullins1949permanent} and plays a crucial role in the observed residual stretches. Here we consistently   
describe such induced anisotropy by means of the classical microsphere model proposed in \cite{goktepe2005micro} in the field of rubber elasticity. In this vein, an isotropic three-dimensional distribution of chains is considered. Moreover we introduce the constraint arising from the network and the resulting entropy variation by considering a new energy term due to the filament interaction with surrounding forest chains. Finally, our approach takes into account the relevant unfolding $\rightleftarrows$ refolding phenomena of crystalline regions composing biodegradable copolymers. As a consequence our approach is able to describe hysteretic behaviors, permanent damage (Mullins effect \cite{mullins1969softening,diani2009review}) and induced anisotropy. We remark that our model bridges the micro-to-macro scale by means of the classical affinity hypothesis diffusely adopted  in rubberlike elasticity. 

We argue that our constitutive model can be inspiring in the perspective of engineering design of new biomedical materials, tailored for innovative technological applications.

As a particular case, we show the effectiveness of our model by considering the experimental results on uniaxial cyclic tests that we performed on a biodegradable copolymer (poliglecaprone 25) used for suture threads.
The model, describing all the effects reported above, depends upon only five microscopically-based parameters with a clear physical interpretation. Interestingly this feature of the model helps us determining a reasonable domain of parameters variation in view of the identification procedure. To this end we carry out a standard optimization procedure (a hybrid particle swarm method refined by a standard local minimization). The model exhibits a very efficient capacity in reproducing experimental results with a global error of 2.62\%. 

While to describe in details all the obtained results in this paper we restricted our attention to poliglecaprone 25 sutures, we remark the present model generality will be useful in describing the mechanical response of polymers with similar microstructure of interest in material science and biomedical applications. Such extensions will be the subject of our future studies.

\section*{Acknowledgments}
G. Vitucci is supported by the POR Puglia FESR-FSE project REFIN A1004.22, F. Trentadue and D. De Tommasi by the Prin Project 2017J4EAYB, G. Puglisi by the Prin Project 2017KL4EF3 and Gruppo Nazionale per la Fisica Matematica (GNFM) of the Istituto Nazionale di Alta Matematica (INdAM).  

\vspace{2cm}
\bibliography{biblio} 

\begin{thebibliography}{10}

\bibitem{trentadue2021predictive}
F.~Trentadue, D.~De~Tommasi, and G.~Puglisi, ``A predictive
  micromechanically-based model for damage and permanent deformations in
  copolymer sutures,'' {\em Journal of the Mechanical Behavior of Biomedical
  Materials}, vol.~115, p.~104277, 2021.

\bibitem{rey2013influence}
T.~Rey, G.~Chagnon, J.-B. Le~Cam, and D.~Favier, ``Influence of the temperature
  on the mechanical behaviour of filled and unfilled silicone rubbers,'' {\em
  Polymer Testing}, vol.~32, no.~3, pp.~492--501, 2013.

\bibitem{martinez2013mechanisms}
J.~S. Martinez, J.-B. Le~Cam, X.~Balandraud, E.~Toussaint, and J.~Caillard,
  ``Mechanisms of deformation in crystallizable natural rubber. part 1: thermal
  characterization,'' {\em Polymer}, vol.~54, no.~11, pp.~2717--2726, 2013.

\bibitem{doi1988theory}
M.~Doi, S.~F. Edwards, and S.~F. Edwards, {\em The theory of polymer dynamics},
  vol.~73.
\newblock oxford university press, 1988.

\bibitem{goktepe2005micro}
S.~G{\"o}ktepe and C.~Miehe, ``A micro--macro approach to rubber-like
  materials. part iii: The micro-sphere model of anisotropic mullins-type
  damage,'' {\em Journal of the Mechanics and Physics of Solids}, vol.~53,
  no.~10, pp.~2259--2283, 2005.

\bibitem{bavzant1986efficient}
P.~Ba{\v{z}}ant and B.~Oh, ``Efficient numerical integration on the surface of
  a sphere,'' {\em ZAMM-Journal of Applied Mathematics and
  Mechanics/Zeitschrift f{\"u}r Angewandte Mathematik und Mechanik}, vol.~66,
  no.~1, pp.~37--49, 1986.

\bibitem{bezwada1995monocryl}
R.~S. Bezwada, D.~D. Jamiolkowski, I.-Y. Lee, V.~Agarwal, J.~Persivale,
  S.~Trenka-Benthin, M.~Erneta, J.~Suryadevara, A.~Yang, and S.~Liu,
  ``Monocryl{\textregistered} suture, a new ultra-pliable absorbable
  monofilament suture,'' {\em Biomaterials}, vol.~16, no.~15, pp.~1141--1148,
  1995.

\bibitem{elias2013stress}
A.~El{\'\i}as-Z{\'u}{\~n}iga, B.~Montoya, W.~Ortega-Lara, E.~Flores-Villalba,
  C.~A. Rodr{\'\i}guez, H.~R. Siller, J.~A. D{\'\i}az-Elizondo, and
  O.~Mart{\'\i}nez-Romero, ``Stress-softening and residual strain effects in
  suture materials,'' {\em Advances in Materials Science and Engineering},
  vol.~2013, 2013.

\bibitem{elias2014rule}
A.~El{\'\i}as-Z{\'u}{\~n}iga, K.~Bayl{\'o}n, I.~Ferrer, L.~Seren{\'o}, M.~L.
  Garcia-Romeu, I.~Bagudanch, J.~Grabalosa, T.~P{\'e}rez-Recio,
  O.~Mart{\'\i}nez-Romero, W.~Ortega-Lara, {\em et~al.}, ``On the rule of
  mixtures for predicting stress-softening and residual strain effects in
  biological tissues and biocompatible materials,'' {\em Materials}, vol.~7,
  no.~1, pp.~441--456, 2014.

\bibitem{rubinstein2003polymer}
M.~Rubinstein, R.~H. Colby, {\em et~al.}, {\em Polymer physics}, vol.~23.
\newblock Oxford university press New York, 2003.

\bibitem{miehe2004micro}
C.~Miehe, S.~G{\"o}ktepe, and F.~Lulei, ``A micro-macro approach to rubber-like
  materials—part i: the non-affine micro-sphere model of rubber elasticity,''
  {\em Journal of the Mechanics and Physics of Solids}, vol.~52, no.~11,
  pp.~2617--2660, 2004.

\bibitem{miehe2005micro}
C.~Miehe and S.~G{\"o}ktepe, ``A micro--macro approach to rubber-like
  materials. part ii: The micro-sphere model of finite rubber
  viscoelasticity,'' {\em Journal of the Mechanics and Physics of Solids},
  vol.~53, no.~10, pp.~2231--2258, 2005.

\bibitem{de2013energetic}
D.~De~Tommasi, N.~Millardi, G.~Puglisi, and G.~Saccomandi, ``An energetic model
  for macromolecules unfolding in stretching experiments,'' {\em Journal of The
  Royal Society Interface}, vol.~10, no.~88, p.~20130651, 2013.

\bibitem{de2015multiscale}
D.~De~Tommasi, G.~Puglisi, and G.~Saccomandi, ``Multiscale mechanics of
  macromolecular materials with unfolding domains,'' {\em Journal of the
  Mechanics and Physics of Solids}, vol.~78, pp.~154--172, 2015.

\bibitem{puglisi2017mic}
G.~Puglisi, D.~De~Tommasi, M.~Pantano, N.~Pugno, and G.~Saccomandi,
  ``Micromechanical model for protein materials: From macromolecules to
  macroscopic fibers,'' {\em Physical Review E}, vol.~96, no.~4, p.~042407,
  2017.

\bibitem{de2010damage}
D.~De~Tommasi, G.~Puglisi, and G.~Saccomandi, ``Damage, self-healing, and
  hysteresis in spider silks,'' {\em Biophysical journal}, vol.~98, no.~9,
  pp.~1941--1948, 2010.

\bibitem{qi2005stress}
H.~J. Qi and M.~C. Boyce, ``Stress--strain behavior of thermoplastic
  polyurethanes,'' {\em Mechanics of materials}, vol.~37, no.~8, pp.~817--839,
  2005.

\bibitem{yeh2003situ}
F.~Yeh, B.~S. Hsiao, B.~B. Sauer, S.~Michel, and H.~W. Siesler, ``In-situ
  studies of structure development during deformation of a segmented poly
  (urethane- urea) elastomer,'' {\em Macromolecules}, vol.~36, no.~6,
  pp.~1940--1954, 2003.

\bibitem{marckmann2002theory}
G.~Marckmann, E.~Verron, L.~Gornet, G.~Chagnon, P.~Charrier, and P.~Fort, ``A
  theory of network alteration for the mullins effect,'' {\em Journal of the
  Mechanics and Physics of Solids}, vol.~50, no.~9, pp.~2011--2028, 2002.

\bibitem{dorfmann2004constitutive}
A.~Dorfmann and R.~W. Ogden, ``A constitutive model for the mullins effect with
  permanent set in particle-reinforced rubber,'' {\em International Journal of
  Solids and Structures}, vol.~41, no.~7, pp.~1855--1878, 2004.

\bibitem{diani2006damage}
J.~Diani, M.~Brieu, and J.~Vacherand, ``A damage directional constitutive model
  for mullins effect with permanent set and induced anisotropy,'' {\em European
  Journal of Mechanics-A/Solids}, vol.~25, no.~3, pp.~483--496, 2006.

\bibitem{machado2012induced}
G.~Machado, G.~Chagnon, and D.~Favier, ``Induced anisotropy by the mullins
  effect in filled silicone rubber,'' {\em Mechanics of Materials}, vol.~50,
  pp.~70--80, 2012.

\bibitem{kuhn1942beziehungen}
W.~Kuhn and F.~Gr{\"u}n, ``Beziehungen zwischen elastischen konstanten und
  dehnungsdoppelbrechung hochelastischer stoffe,'' {\em Kolloid-Zeitschrift},
  vol.~101, no.~3, pp.~248--271, 1942.

\bibitem{fantner2006sacrificial}
G.~E. Fantner, E.~Oroudjev, G.~Schitter, L.~S. Golde, P.~Thurner, M.~M. Finch,
  P.~Turner, T.~Gutsmann, D.~E. Morse, H.~Hansma, {\em et~al.}, ``Sacrificial
  bonds and hidden length: unraveling molecular mesostructures in tough
  materials,'' {\em Biophysical journal}, vol.~90, no.~4, pp.~1411--1418, 2006.

\bibitem{marko1995stretching}
J.~F. Marko and E.~D. Siggia, ``Stretching dna,'' {\em Macromolecules},
  vol.~28, no.~26, pp.~8759--8770, 1995.

\bibitem{col67}
B.~D. Coleman and M.~E. Gurtin, ``Thermodynamics with internal state
  variables,'' {\em The journal of chemical physics}, vol.~47, no.~2,
  pp.~597--613, 1967.

\bibitem{bertotti1998hysteresis}
G.~Bertotti, {\em Hysteresis in magnetism: for physicists, materials
  scientists, and engineers}.
\newblock Gulf Professional Publishing, 1998.

\bibitem{mullins1949permanent}
L.~Mullins, ``Permanent set in vulcanized rubber,'' {\em Rubber Chemistry and
  Technology}, vol.~22, no.~4, pp.~1036--1044, 1949.

\bibitem{mullins1969softening}
L.~Mullins, ``Softening of rubber by deformation,'' {\em Rubber chemistry and
  technology}, vol.~42, no.~1, pp.~339--362, 1969.

\bibitem{diani2009review}
J.~Diani, B.~Fayolle, and P.~Gilormini, ``A review on the mullins effect,''
  {\em European Polymer Journal}, vol.~45, no.~3, pp.~601--612, 2009.

\end{thebibliography}
\bibliographystyle{ieeetr}
\newpage
\appendix
\section{Comparison with existing experimental tests}\label{app:zu}
\begin{figure*}[ht!]
	\centering
	\includegraphics[width=.95\textwidth]{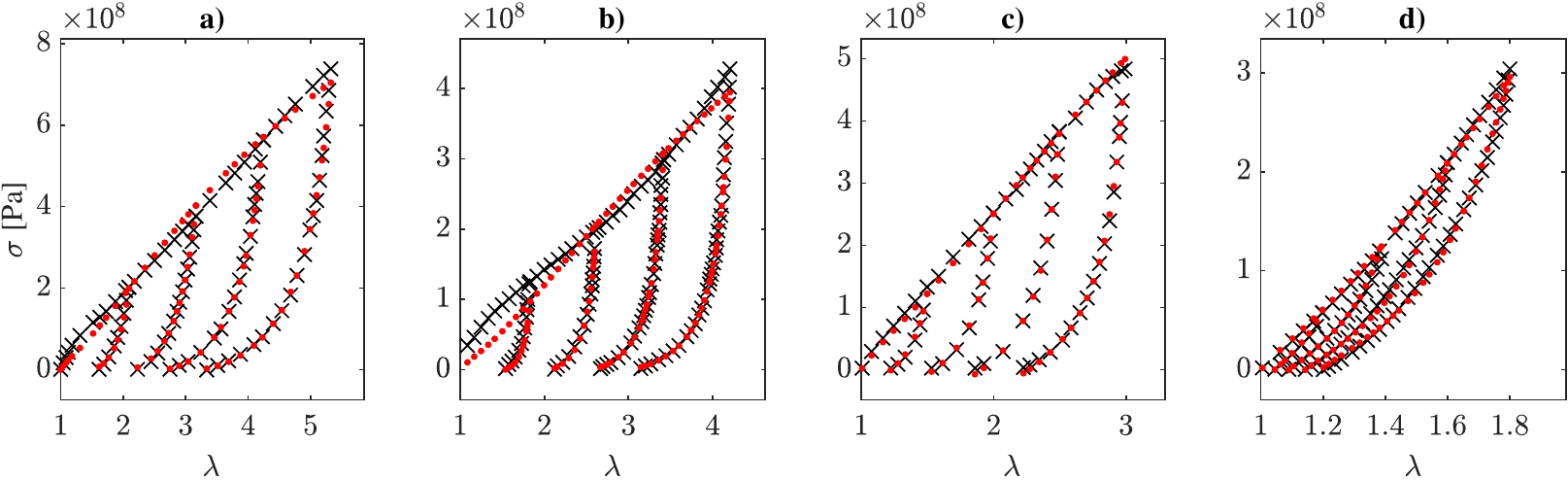}
	\caption{Comparison between modified theory and literature experiments. The evolution law includes the parameter $\beta_{u1}$ defined in \eqrf{eq:lacevolb1}. Uniaxial cyclic tests on four absorbable suture threads carried out by \cite{elias2013stress}. Data are shown in black, theory including \eqrf{eq:lacevolb1} is shown in red. Materials: a) Monoplus; b) Monosyn; c) Polydioxanone 2.0; d)  Polydioxanone 4.0.}
	\label{fig:4x}
\end{figure*}
\begin{table*}[th!]\centering
	\begin{tabular}{ | c| c| c| c| c| c| c| }
		\hline 
		&$\kappa$[$10^8\times$Pa]&$\nu$[$10^8\times$Pa]&$\laco$&$\beta_{u0} $&$\beta_{u1} $&err[\%] \\
		\hline  
		Monoplus&3.421&1.499&1.339&-0.012&-0.091&3.94\\
		\hline
		Monosyn&1.023& -0.130&1.201&-0.986&0.068&6.77\\ \hline
		Polydioxanone 2.0&3.450&-1.298 & 1.3107&0.557&-0.688 &2.07\\ \hline
		Polydioxanone 4.0& 0.453&-6.010&1.064&-0.232&0.395&3.04 \\
		\hline
	\end{tabular}\caption{Parameters of the fit with $\beta_{u1}$ defined in \eqrf{eq:lacevolb1}.}
	\label{tab:pfitallb1}
\end{table*}
In this Section we briefly explore the performance of our model on four other absorbable suture threads experimental data as published by \cite{elias2013stress}. In order to adapt our formulation to this larger class of materials, we introduce an additional parameter $\beta_{u1}$ as follows:
\begin{equation}\label{eq:lacevolb1}	
	\lambda_n=\sqrt{\lamax}\left(\frac{\lambda}{\lamax}\right)^{{\beta_{u}+\beta_{u1}\lambda+\beta_r\sqrt{\lambda-\lamin}}}\geq\bar{\lambda}_n\quad.
\end{equation}
The results of the parameter identification procedure are shown in Fig.\ref{tab:pfitallb1} and Table \ref{tab:pfitallb1}. Notice that the lack of reloading data does not allow estimating the reloading parameter $\beta_r$ so that for this set of experiments we only need five parameters even including $\beta_{u1}$. 

Notice that the model of \cite{elias2013stress,elias2014rule} required seven parameters, it did not include internal hysteresis and did not produce a better match than the present work. 

The errors of the fit result extremely low and this makes us confident that the developed multiscale model is physically sound and versatile in the perspective of describing the mechanical behavior of a vast class of polymeric materials of biomedical interest. Further analysis will be the subject of future work.

\end{document}